\documentclass{emulateapj}

\usepackage{lscape}

\newcommand{\about}{$\sim\!\!$~}

\newcommand{\be}{\begin{displaymath}}
\newcommand{\ee}{\end{displaymath}}

\def\lsim{\hbox{\rlap{\raise 0.425ex\hbox{$<$}}\lower 0.65ex\hbox{$\sim$}}}
\def\gsim{\hbox{\rlap{\raise 0.425ex\hbox{$>$}}\lower 0.65ex\hbox{$\sim$}}}

\newcommand{\kms}{km~s$^{-1}$ }

\newcommand{\dof}{\rm dof}

\shorttitle{ESSENCE Spectroscopy: The First 4 Years}
\shortauthors{Foley et~al.}

\begin{document}

\title{Spectroscopy of High-Redshift Supernovae from the ESSENCE Project: The First Four Years}

\author{R.~J.~Foley\altaffilmark{1,2,3}, %
{T.~Matheson}\altaffilmark{4}, %
{S.~Blondin}\altaffilmark{2,5}, %
{R.~Chornock}\altaffilmark{1}, %
{J.~M.~Silverman}\altaffilmark{1}, %
{P.~Challis}\altaffilmark{2}, %
{A.~Clocchiatti}\altaffilmark{6}, %
{A.~V.~Filippenko}\altaffilmark{1}, %
{R.~P.~Kirshner}\altaffilmark{2}, %
{B.~Leibundgut}\altaffilmark{5}, %
{J.~Sollerman}\altaffilmark{7,8}, %
{J.~Spyromilio}\altaffilmark{5}, %
{J.~L.~Tonry}\altaffilmark{9}, 	
{T.~M.~Davis}\altaffilmark{10,7}, %
{P.~M.~Garnavich}\altaffilmark{11}, %
{S.~W.~Jha}\altaffilmark{12,1,13}, %
{K.~Krisciunas}\altaffilmark{14}, %
{W.~Li}\altaffilmark{1}, %
{G.~Pignata}\altaffilmark{15}, %
{A.~Rest}\altaffilmark{16,17}, %
{A.~G.~Riess}\altaffilmark{18,19}, %
{B.~P.~Schmidt}\altaffilmark{20}, %
{R.~C.~Smith}\altaffilmark{17}, %
{C.~W.~Stubbs}\altaffilmark{2,16}, %
{B.~E.~Tucker}\altaffilmark{20}, and %
{W.~M.~Wood-Vasey}\altaffilmark{2,21}
}
\email{rfoley@cfa.harvard.edu}

\altaffiltext{1}{Department of Astronomy, University of California, Berkeley, CA 94720-3411.}
\altaffiltext{2}{Harvard-Smithsonian Center for Astrophysics, 60 Garden Street, Cambridge, MA 02138.}
\altaffiltext{3}{Clay Fellow.}
\altaffiltext{4}{National Optical Astronomy Observatory, 950 North Cherry Avenue, Tucson, AZ 85719-4933.}
\altaffiltext{5}{European Southern Observatory, Karl-Schwarzschild-Strasse 2, D-85748 Garching, Germany.}
\altaffiltext{6}{Pontificia Universidad Cat\'olica de Chile, Departamento de Astronom\'ia y Astrof\'isica, Casilla 306, Santiago 22, Chile.}
\altaffiltext{7}{Dark Cosmology Centre, Niels Bohr Institute, University of Copenhagen, Juliane Maries Vej 30, DK-2100 Copenhagen \O, Denmark.}
\altaffiltext{8}{Department of Astronomy, Stockholm University, AlbaNova, 10691 Stockholm, Sweden.}
\altaffiltext{9}{Institute for Astronomy, University of Hawaii, 2680 Woodlawn Drive, Honolulu, HI 96822.}
\altaffiltext{10}{Department of Physics, University of Queensland, QLD, Australia 4072.}
\altaffiltext{11}{Department of Physics, University of Notre Dame, 225 Nieuwland Science Hall, Notre Dame, IN 46556-5670.}
\altaffiltext{12}{Department of Physics and Astronomy, Rutgers, the State University of New Jersey, 136 Frelinghuysen Road, Piscataway, NJ 08854.}
\altaffiltext{13}{Kavli Institute for Particle Astrophysics and Cosmology, Stanford Linear Accelerator Center, 2575 Sand Hill Road, MS 29, Menlo Park, CA 94025.}
\altaffiltext{14}{Department of Physics, Texas A\&M University, College Station, TX 77843-4242.}
\altaffiltext{15}{Departamento de Astronomia, Universidad de Chile, Casilla 36-D, Santiago, Chile.}
\altaffiltext{16}{Department of Physics, Harvard University, 17 Oxford Street, Cambridge, MA 02138.}
\altaffiltext{17}{National Optical Astronomy Observatory / Cerro Tololo Inter-American Observatory, Casilla 603, La Serena, Chile.}
\altaffiltext{18}{Space Telescope Science Institute, 3700 San Martin Drive, Baltimore, MD 21218.}
\altaffiltext{19}{Johns Hopkins University, 3400 North Charles Street, Baltimore, MD 21218.}
\altaffiltext{20}{The Research School of Astronomy and Astrophysics, The Australian National University, Mount Stromlo and Siding Spring Observatories, via Cotter Road, Weston Creek, PO 2611, Australia.}
\altaffiltext{21}{Department of Physics and Astronomy, University of Pittsburgh, 100 Allen Hall, Pittsburgh, PA 15260.}

\begin{abstract}
We present the results of spectroscopic observations from the ESSENCE
high-redshift supernova (SN) survey during its first four years of
operation.  This sample includes spectra of all SNe~Ia whose light
curves were presented by \citet{Miknaitis07} and used in the
cosmological analyses of \citet{Davis07} and \citet{Wood-Vasey07}.
The sample represents 273 hours of spectroscopic observations with
6.5--10-m-class telescopes of objects detected and selected for
spectroscopy by the ESSENCE team. We present 174 spectra of 156
objects.  Combining this sample with that of \citet{Matheson05}, we
have a total sample of 329 spectra of 274 objects.  From this, we are
able to spectroscopically classify 118 Type Ia SNe.  As the survey has
matured, the efficiency of classifying SNe~Ia has remained constant
while we have observed both higher-redshift SNe~Ia and SNe~Ia farther
from maximum brightness.  Examining the subsample of SNe~Ia with
host-galaxy redshifts shows that redshifts derived from only the SN~Ia
spectra are consistent with redshifts found from host-galaxy spectra.
Moreover, the phases derived from only the SN~Ia spectra are
consistent with those derived from light-curve fits.  By comparing our
spectra to local templates, we find that the rate of objects similar
to the overluminous SN~1991T and the underluminous SN~1991bg in our sample
are consistent with that of the local sample.  We do note, however,
that we detect no object spectroscopically or photometrically similar
to SN~1991bg.  Although systematic effects could reduce the
high-redshift rate we expect based on the low-redshift surveys, it is
possible that SN~1991bg-like SNe~Ia are less prevalent at high
redshift.
\end{abstract}


\section{Introduction}\label{s:intro}

A decade after the observations of high-redshift Type Ia supernovae
(SNe~Ia) caused a sea change in cosmology, indicating that the
expansion rate of the Universe is currently accelerating
\citep{Riess98:Lambda, Perlmutter99} (see \citealt{Filippenko05} for
a review), we are still far from understanding the nature of the
``dark energy'' that causes this accelerated expansion.  Recent
high-redshift SN~Ia surveys, using the large and normalized  peak
luminosities of SNe~Ia, have focused on measuring the
equation-of-state parameter of the dark energy, $w = P/(\rho c^{2})$,
currently constraining it to \about10\% \citep{Astier06, Riess07,
Wood-Vasey07, Kowalski08}.

The determination of the expansion history of the Universe with SNe~Ia
is performed by measuring the luminosity distance as a function of
redshift.  A relationship between the light-curve shape of a SN~Ia and
its luminosity is used to obtain precise luminosity distances
\citep[e.g.,][]{Phillips93, Jha07, Guy07}.  The redshift of the object
is typically found through spectroscopy of the SN by cross-correlating
with low-redshift template spectra or its host galaxy.  Although
photometry alone can be powerful \citep[e.g.,][]{Barris04,
Poznanski07}, only spectroscopy can currently provide the accurate
classification and redshifts necessary for estimating cosmological
parameters.

Besides providing the redshift for each object, spectroscopy allows
detailed studies of the physics of high-redshift SNe~Ia.  A critical
aspect of high-redshift SN~Ia surveys is to only include SNe~Ia, and
not other transient objects, in their final analysis.  Since the
optical spectra of SNe~Ia are distinct from most transient phenomena,
spectroscopy provides a consistent and precise method for determining
the nature of each transient \citep[see][for a review of SN
spectroscopy]{Filippenko97}.  Furthermore, with high-redshift SN~Ia
spectra one can compare physical properties of the objects with their
low-redshift counterparts \citep[e.g.,][]{Blondin06, Balland07,
Garavini07, Bronder08, Ellis08, Foley08}.

In order to measure $w$ to $\lesssim 10\%$, the Equation of State:
SupErNovae trace Cosmic Expansion (ESSENCE) team has concluded a
six-year NOAO survey project using the CTIO 4~m telescope and the
MOSAIC~II camera with the intention of discovering and following
\about 200 SNe~Ia over the redshift range of $0.2 < z < 0.8$
\citep{Miknaitis07}.  During the first four years of the ESSENCE
survey, which observed approximately October through December, several
hundred transient objects were detected \citep{Miknaitis07}.  As most
of these objects were relatively faint ($R > 21$~mag), a large amount
of telescope time at 6.5--10~m telescopes was required for
spectroscopic follow-up observations. Despite being awarded
approximately 100 nights at these facilities over the first four years
of the survey, we were still not able to obtain a spectrum of every
candidate object \citep{Matheson05}.

However, during the first 12 search months (ESSENCE searched three
months per year), we were able to classify 121 SNe~Ia.  In this paper,
we present the spectra of all spectroscopic targets observed during
the first four years of the ESSENCE survey.  This sample, which
updates and supersedes the analysis of \citet{Matheson05} for the
spectroscopy of the first two years of the ESSENCE survey, includes
all SNe~Ia presented by \citet{Miknaitis07} and used in the analysis
of \citet{Wood-Vasey07} and \citet{Davis07}.  We discuss our
observations in \S~\ref{s:observations}.  The classification scheme
presented by \citet{Matheson05}, \citet{Miknaitis07}, and
\citet{Blondin07} is updated in \S~\ref{s:classification} \citep[the
method is similar to that presented by][]{Miknaitis07}.  In
\S~\ref{s:sample}, we discuss the properties of our sample, and we
summarize our conclusions in \S~\ref{s:conclusions}.

\setcounter{footnote}{0}

\section{Observations}\label{s:observations}

Over the first four years of the ESSENCE survey, we detected several
thousand transient objects.  Since we did not have adequate observing
time at large telescopes to follow each transient spectroscopically,
we prioritized the objects by a combination of likelihood to be a
SN~Ia, observational ease, and, occasionally, likely redshift and
phase relative to maximum light.  These criteria were all determined
from our search images, using features such as color and offset from
the host galaxy.  In particular, the likelihood of an object being a
SN~Ia is deduced from an approximate photo-$z$ of its host, its $R-I$
color \citep{Tonry03}, and its rise time.  Objects which had
particularly blue colors were considered more likely to be
core-collapse SNe, while objects with a slow rise time were more
likely to be active galactic nuclei (AGNs) or other transients.
Further details about the target selection can be found in
\citet{Matheson05} and \citet{Miknaitis07}.

Spectroscopic observations of ESSENCE targets were obtained at a wide
variety of telescopes: the Keck I and II 10~m telescopes, the European
Southern Observatory 8.2~m Very Large Telescope (ESO VLT), the Gemini
North and South 8~m telescopes, the Magellan Baade and Clay 6.5~m
telescopes, and the MMT 6.5~m telescope.  The spectrographs used were
LRIS \citep{Oke95} with Keck I, DEIMOS \citep{Faber03} with Keck II,
FORS1 \citep{Appenzeller98} with the VLT, GMOS \citep{Hook04} with
Gemini (North and South), IMACS \citep{Dressler06} with Baade, LDSS2
\citep{Allington-Smith94} and
LDSS3\footnote{http://www.lco.cl/telescopes-information/magellan/instruments-1/ldss-3-1/
.} with Clay, and the Blue Channel \citep{Schmidt89} with
MMT.  Nod-and-shuffle techniques \citep{Glazebrook01} were used with
GMOS (North and South) and some IMACS observations to improve sky
subtraction in the red portion of the spectrum.  A single spectrum was
obtained with the FAST spectrograph \citep{Fabricant98} mounted on the
Tillinghast 1.5~m telescope at the F.~L.\ Whipple Observatory.

Standard CCD processing and spectrum extraction were accomplished with
IRAF\footnote{IRAF: the Image Reduction and Analysis Facility is
distributed by the National Optical Astronomy Observatory, which is
operated by the Association of Universities for Research in Astronomy,
Inc. (AURA) under cooperative agreement with the National Science
Foundation (NSF).}.  Most of the data were extracted using the optimal
algorithm of \citet{Horne86}; for the ESO VLT data, an alternative
extraction method based on Richardson-Lucy restoration
\citep{Blondin05} was employed.  Low-order polynomial fits to
calibration-lamp spectra were used to establish the wavelength scale.
Small adjustments derived from night-sky lines in the object frames
were applied.  We employed our own IDL routines to flux calibrate the
data and remove telluric lines using the well-exposed continua of the
spectrophotometric standards \citep{Wade88, Foley03}.

The photometric data were obtained on the NOAO Blanco 4~m telescope
with the MOSAIC~II imaging camera.  The light curves for the SNe~Ia
are available online\footnote{See http://www.ctio.noao.edu/essence/}
\citep{Miknaitis07}.

\begin{figure}
\centering
\epsscale{0.8}
\rotatebox{90}{
\plotone{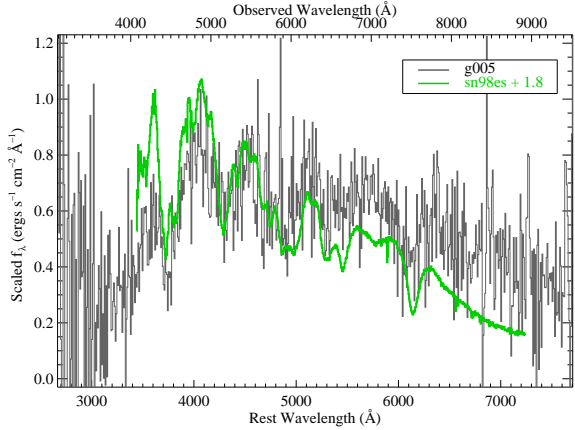}}
\caption{Spectrum of g005 at $z = 0.20$ compared to SN~1998es, a
SN~1991T-like SN~Ia, at $t = 1.8$~d relative to $B$-band maximum.
Differences in the continuum shape are ignored by SNID, which removes a pseudo-continuum from the spectra before cross-correlation.}\label{f:spec1}
\end{figure}

\begin{figure}
\centering
\epsscale{0.8}
\rotatebox{90}{
\plotone{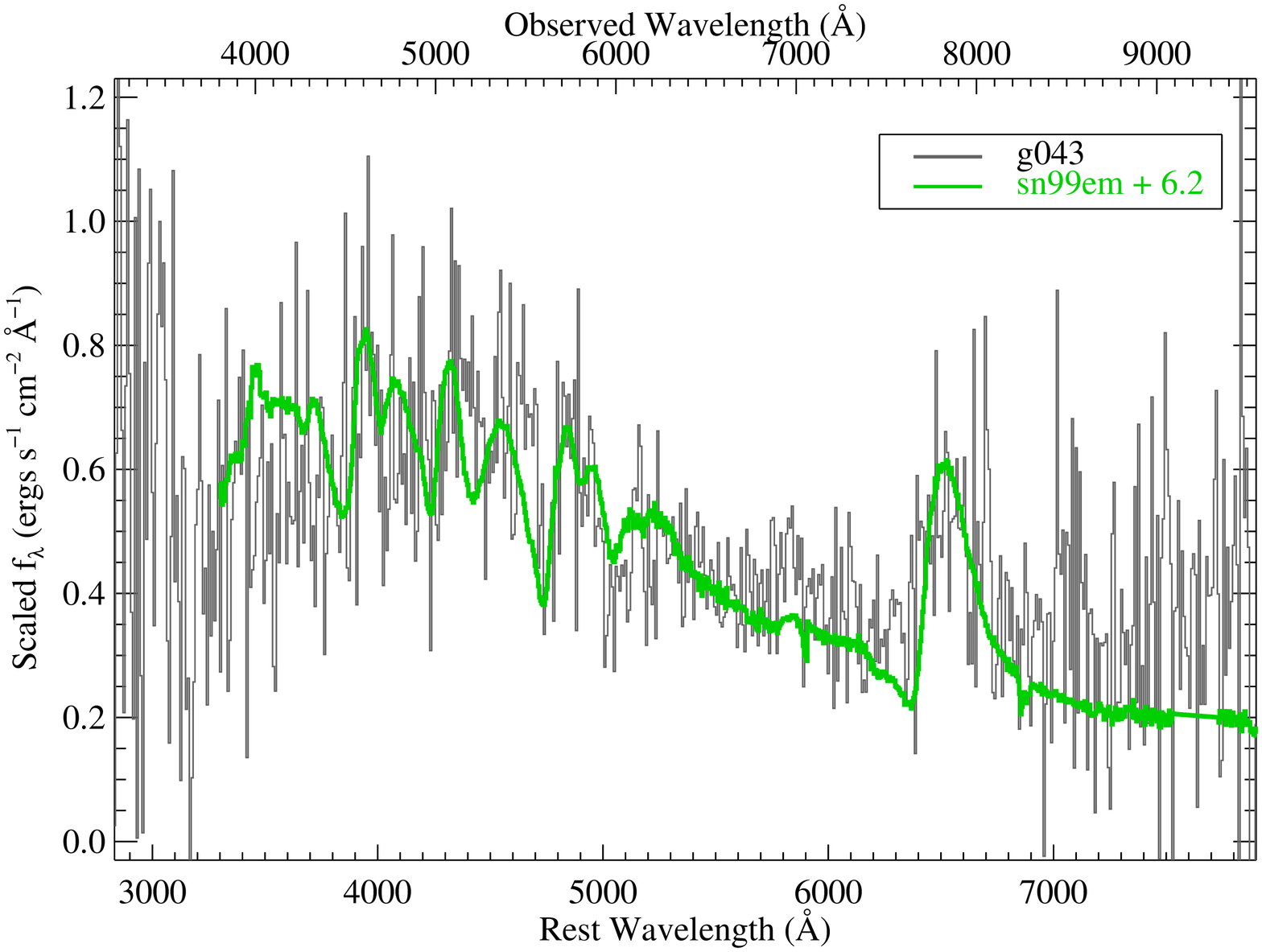}}
\caption{Spectrum of g043 at $z = 0.187$ compared to SN~1999em, a
SN~II-P.  Differences in the continuum shape are ignored by SNID, which
attempts to remove a pseudo-continuum by fitting a low-order
polynomial to the spectra before cross-correlation.}\label{f:spec2}
\end{figure}

\begin{figure}
\centering
\epsscale{0.8}
\rotatebox{90}{
\plotone{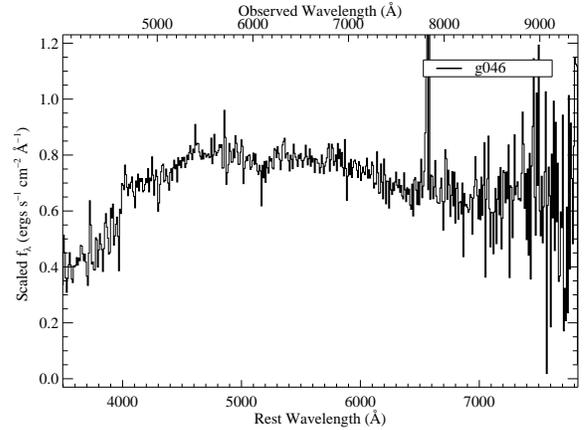}}
\caption{Spectrum of g046 at $z = 0.184$.  The spectrum is dominated
by galaxy light with no detectable amount of light from a transient
object.}\label{f:spec3}
\end{figure}

\begin{figure}
\centering
\epsscale{0.8}
\rotatebox{90}{
\plotone{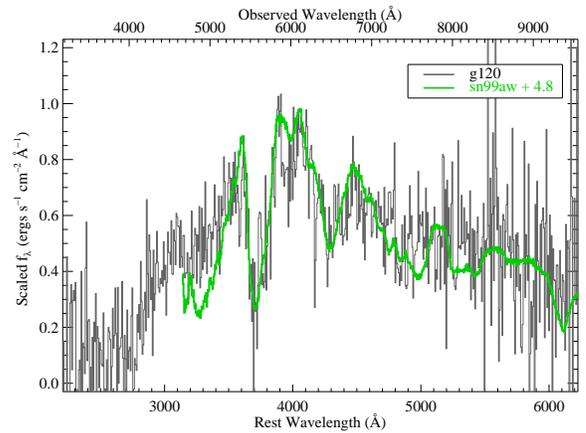}}
\caption{Spectrum of g120 at $z = 0.51$ compared to SN~1999aw, a
SN~Ia, at $t = 4.8$~d relative to $B$ maximum.  We were unable to
determine a subtype for g120.  Differences in the continuum shape are
ignored by SNID, which attempts to remove a pseudo-continuum by
fitting a low-order polynomial to the spectra before
cross-correlation.}\label{f:spec4}
\end{figure}

\begin{figure}
\centering
\epsscale{0.8}
\rotatebox{90}{
\plotone{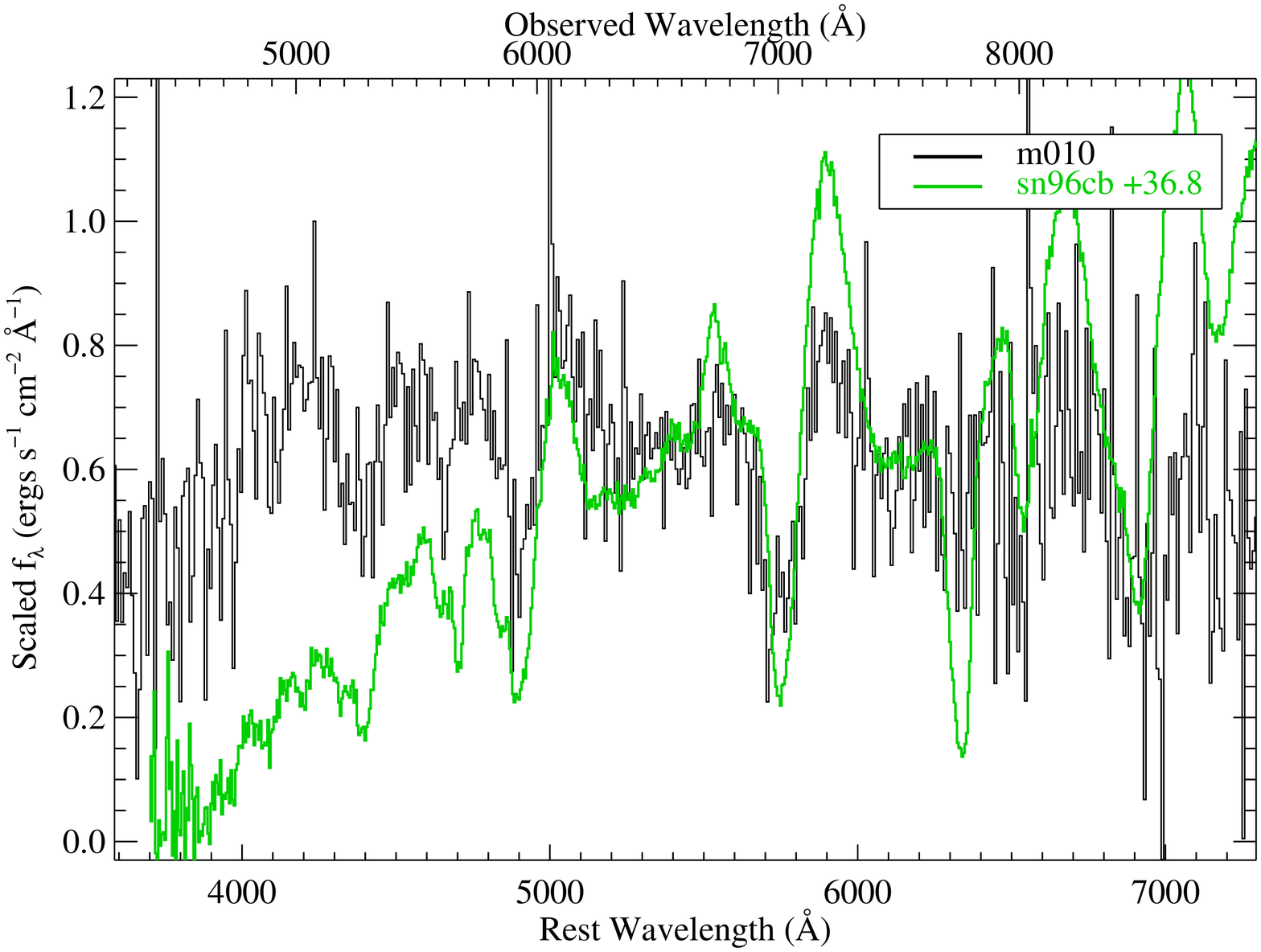}}
\caption{Spectrum of m010 at $z = 0.216$ compared to SN~1996cb, a
SN~IIb.  Differences in the continuum shape are ignored by SNID, which
attempts to remove a pseudo-continuum by fitting a low-order
polynomial to the spectra before cross-correlation.}\label{f:spec5}
\end{figure}

\begin{figure}
\centering
\epsscale{0.8}
\rotatebox{90}{
\plotone{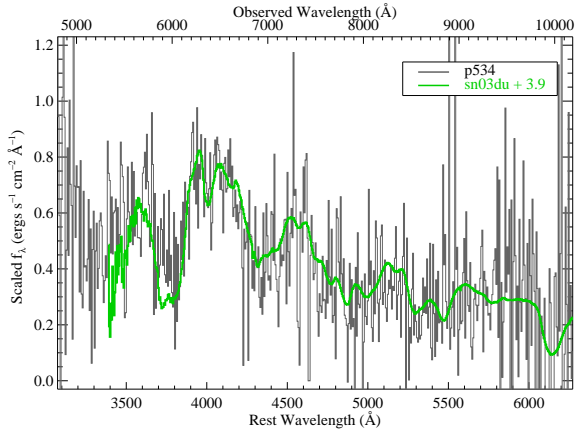}}
\caption{Spectrum of p534 at $z = 0.621$ after galaxy subtraction from
the \texttt{superfit} routine compared to SN~2003du, a SN~Ia.
Differences in the continuum shape are ignored by SNID, which removes
a pseudo-continuum from the spectra before cross-correlation.
Additional figures for all objects listed in Table~\ref{t:sntable} can
be found in the online edition.}\label{f:spec6}
\end{figure}

In Table~\ref{t:sntable}, we present a full list of our observations
(date of observation, telescope/instrument, and exposure times).  We
also include information about the nature of each object (object
classification, redshift, and, if a SN~Ia, phase and light-curve
shape).  We present the spectra from 2004 and 2005 in
Figures~\ref{f:spec1}--\ref{f:spec6} ({\it All additional spectra are
available in the online edition}).  For each SN~Ia, we plot both the
spectrum of the SN as well as the best-fit SNID \citep{Blondin07}
template spectrum.

\section{Object Classification}\label{s:classification}
As all cosmological results resulting from the ESSENCE survey depend
on having an uncontaminated sample of SNe~Ia, we must pay particular
attention to proper object classification.  SN classification is based
on the optical spectrum \citep{Filippenko97, Turatto03}.  Type II SNe
are characterized by the presence of obvious hydrogen lines, while
Type I SNe lack hydrogen.  Type I SNe are further distinguished by the
presence of He (Type Ib), strong \ion{Si}{2} $\lambda$6355 (Type Ia), or the 
lack (or weak presence) of both (Type Ic). The velocity of the SN ejecta 
causes the features to be blueshifted, with the \ion{Si}{2} $\lambda$6355
feature typically being observed in absorption at \about 6150~\AA.

With fully calibrated spectra, we can attempt to classify the object
by physical origin.  SNe, having broad (\about 10,000~\kms) spectral
features, are very distinct from AGNs, galaxies, stars, and other
astrophysical objects.  However, it can occasionally be difficult to
distinguish among SN types.  High-redshift SN spectra typically have low 
signal-to-noise ratio (S/N) and considerable host-galaxy contamination, 
while lacking the \ion{Si}{2} $\lambda$6355 feature (which is redshifted
outside the optical range), which can make differentiating between SN
types quite difficult.  Accordingly, we have implemented the SNID
algorithm \citep{Blondin07} to determine SN types.

SNID correlates an input spectrum with a large database of template
spectra at zero redshift.  These include nearby ($z<0.1$) SNe of all
types (Ia, Ib, Ic, II), as well as galaxies, AGNs, luminous blue
variables (LBVs), and other variable stars \citep[see][for the complete
database listing]{Blondin07}.  The SNID algorithm has been extensively
described by \citet{Blondin07} and already used to determine the type,
redshift, and age of ESSENCE SN spectra by \citet{Matheson05} and
\citet{Miknaitis07}, and we refer the reader to these papers for more
details (in particular for what is meant by a ``good'' correlation).

We also attempt to further divide each SN type into subtypes as
follows: Ia --- Ia-norm, Ia-pec, Ia-91T, Ia-91bg; Ib --- Ib-norm,
Ib-pec, IIb; Ic --- Ic-norm, Ic-pec, Ic-broad; II --- IIP, II-87A,
IIL, IIn, IIb. ``Norm'' and ``pec'' subtypes are used to identify the
spectroscopically ``normal'' and ``peculiar'' SNe of a given type
\citep[see][for a detailed description of each subtype]{Blondin07}.
For SNe~Ia, ``91T'' indicates spectra that resemble those of the
overluminous SN~1991T \citep{Filippenko92:91T, Phillips92}, and also
includes SN~1999aa-like objects \citep{Strolger02, Garavini04};
``91bg'' indicates spectra that resemble those of the subluminous
SN~1991bg \citep{Filippenko92:91bg, Leibundgut93}, and includes all
subluminous objects characterized by a stronger Si~II~$\lambda$5800
line \citep[e.g., SN~1999gh;][]{Matheson08}.  The spectra that
correspond to the ``Ia-pec'' category in this case are those of
SN~2000cx-like \citep{Li01:00cx} and SN~2002cx-like \citep{Li03} events.  
These classifications correspond roughly to the categories defined by
\citet{Branch93:normal}.  For Type
Ic SNe, ``Ic-broad'' is used to identify broad-lined SNe~Ic (often
referred to as ``hypernovae'' in the literature; see
\citealt{Galama98} and \citealt{Foley03} for well-known examples),
some of which are associated with gamma-ray bursts.  The notation
adopted for the Type II subtypes is commonly used in the literature.
Note that Type IIb SNe (whose spectra evolve from a Type II to a Type
Ib, as in SNe~1987K and 1993J --- see \citealt{Filippenko88}, 
\citealt {Filippenko93}, and \citealt{Matheson00:93j}) are included
both in the ``Ib'' and ``II'' types.  All non-SN templates are grouped
in the ``NotSN'' type.  We are unable to determine a specific subtype
for many objects.

We classify the input spectra in a similar manner to that outlined by
\citet{Miknaitis07}.  More specifically, we execute four SNID runs
to separately determine the type, subtype, redshift, and age of the
input spectrum, as follows.

\begin{itemize}

\item[1.]{{\bf Type:} The input spectrum is asserted to be of a given
type when the fraction of ``good'' correlations that correspond to
this type exceeds 50\%.  In addition, we require the best-match SN
template to be of the same type.  When the redshift is known
beforehand (from narrow emission or absorption lines associated with
the host galaxy), we force SNID to only look for correlations at this
redshift ($\pm0.02$).  If SNID determines the type of the input
spectrum, an attempt is made to determine its subtype.}

\item[2.]{{\bf Subtype:} To determine the subtype we only consider
``good'' correlations with template spectra corresponding to the
previously determined type.  The input spectrum is asserted to be of a
given subtype when the fraction of ``good'' correlations that
correspond to this subtype exceeds 50\%.  In addition, we require the
best-match SN template to be of the same subtype.  Examples of SNe~Ia
for which we could and could not determine a subtype are g005
(displayed in Figure~\ref{f:spec1}) and g120 (displayed in
Figure~\ref{f:spec4}, respectively.  Again, we use the galaxy redshift
when available.  Regardless of whether SNID determines a subtype, a third run
is executed to determine the redshift.}

\item[3.]{{\bf Redshift:} The SNID redshift corresponds to the median
of all ``good'' template redshifts, while the redshift error is given
as the standard deviation of these same redshifts \citep[see][for more
details]{Blondin07}.  If a subtype has been determined, we only
consider ``good'' correlations with template spectra corresponding to
this subtype; otherwise, all template spectra of the given type are
used.  Here no prior information on the galaxy redshift is used. If a
redshift is determined in this run, a fourth run is executed to
determine the age.}

\item[4.]{{\bf Age:} The SNID age corresponds to the median of all
``good'' template ages, while the age error is given as the standard
deviation of these same ages.  If a subtype has been determined, we
only consider ``good'' correlations with template spectra corresponding to
this subtype; otherwise, all template spectra corresponding to the
given type are used (as for the redshift determination).  The redshift
is fixed to the host-galaxy redshift when available, or to the
previously determined SNID redshift otherwise ($\pm0.02$).
\citet{Blondin07} mention that the age error calculated this way is
typically overestimated, but here we make no attempts to compute a
more robust age error (as done by \citealt{Blondin08}).}

\end{itemize}

Note that we have run SNID on {\it all} ESSENCE spectra, but only
report the SNID output for spectra classified as SNe
(Table~\ref{t:sntable}). For all other spectra, we rely on the visual
classification by spectroscopy experts within our team.  In a few
cases, a spectrum not classified as a SN~Ia by SNID is found to be
consistent with a SN~Ia spectrum through visual inspection.  For these
cases we report a ``Ia?'' classification.  In fact, SNID can sometimes
fail to correctly classify an input SN~Ia spectrum, mainly because of
low S/N or excessive galaxy contamination.  \citet{Blondin07} showed
that the redshift and age determination with SNID is greatly affected
for input spectra that consist of more than 50\% galaxy light.

\begin{figure}
\centering
\epsscale{1.5}
\rotatebox{90}{
\plotone{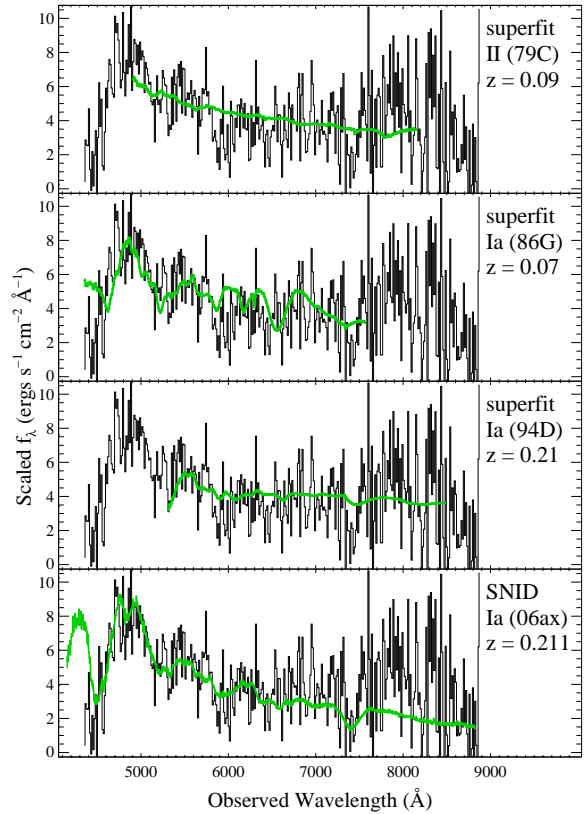}}
\caption{Observed spectrum of ESSENCE SN n404.  The green lines are
low-redshift comparison spectra.  The top three panels show the first
(SN~II~1979C at $z = 0.09$), fourth (first SN~Ia; SN~1986G at $z =
0.07$), and eighth (first SN~Ia at $z \approx 0.21$; SN~1994D)
best-matched low-redshift template spectra from \texttt{superfit} in
green.  The bottom panel shows the best-matched low-redshift template
spectrum from SNID in green.  Differences in the continuum shape are
ignored by SNID, which removes a pseudo-continuum from the spectra
before cross-correlation; therefore, the continuum of the SNID
template may not exactly match that of n404.}\label{f:snid_superfit}
\end{figure}

In addition to using SNID, we have applied an alternate technique for
SN classification that is based on a $\chi^{2}$ minimization of
fitting a SN spectrum to template spectra with a variable extinction
and galaxy contamination \citep{Howell05}.  This method is implemented
in the \texttt{superfit} IDL package.  The main advantage of the
$\chi^{2}$ minimization technique is that it can classify SNe from
spectra that are highly contaminated by galaxy light.  One of the main
drawbacks of \texttt{superfit} is that it assumes that the continuum of
a high-redshift SN~Ia must be similar to that of a low-redshift
SN~Ia.  Although usually this can be worked around by having a
different value for the extinction (which will change the continuum
shape of the spectrum even if the value of the ``extinction'' is not
strictly physical) or a different galaxy template, there is the
possibility that \texttt{superfit} will not classify some objects that
SNID does.  This is demonstrated by n404 in
Figure~\ref{f:snid_superfit}.  For this object, SNID classifies it as
a SN~Ia at $z = 0.211$.  For the same object, \texttt{superfit}
classifies it as a SN~II, with 3 of the top 5 matches with
low-redshift template spectra being SNe~II.  The best-fit SN~Ia
template places n404 at z = 0.07.  One has to go to the eighth-best
match to obtain a SN~Ia at the same redshift as SNID.  Having
performed a ``brute-force,'' where we did not carefully adjust
various parameters to achieve the best results for each spectrum,
classifcation for all spectra in our sample, we found other object in
our sample where SNID and \texttt{superfit} differ.  However, a more
careful \texttt{superfit} analysis may have very few discrepancies
with SNID.  The other main drawback of \texttt{superfit} is that
although \texttt{superfit} produces a $\chi^{2}$-like goodness-of-fit
parameter, there is no formal evaluation of the uncertainty of any of
the derived values (cf, the redshift error in SNID is related to the
height of the correlation peak; see \citealt{Blondin07}).

Applying the $\chi^{2}$ minimization technique to all spectra which
were not originally classified by SNID, we obtained several candidate
SNe.  The \texttt{superfit} package will output a galaxy-subtracted
spectrum using a galaxy fraction (and galaxy type) that best matches a
given SN template spectrum.  Since we have not modeled the robustness
of the goodness-of-fit parameter from this technique, we attempted to
classify the galaxy-subtracted spectra with SNID.  We fit the top five
galaxy-subtracted \texttt{superfit} spectra with SNID, looking for a
consensus result.  For 99 potential candidates, this process yielded
one additional SN~Ia? (f123), as well as identifying a second-epoch
spectrum of a SN~Ia (p534; as identified from the other spectrum by
SNID) as a SN~Ia.

There is some concern that extra degrees of freedom are
introduced into our SNID fitting as the \texttt{superfit} packages
changes the spectrum by applying an extinction estimate and removing a
galaxy spectrum from each input spectrum.  However, since SNID
normalizes all spectra to have a flat continuum, it is relatively
insensitive to reddening.

\begin{figure}
\centering
\epsscale{1.1}
\rotatebox{0}{
\plotone{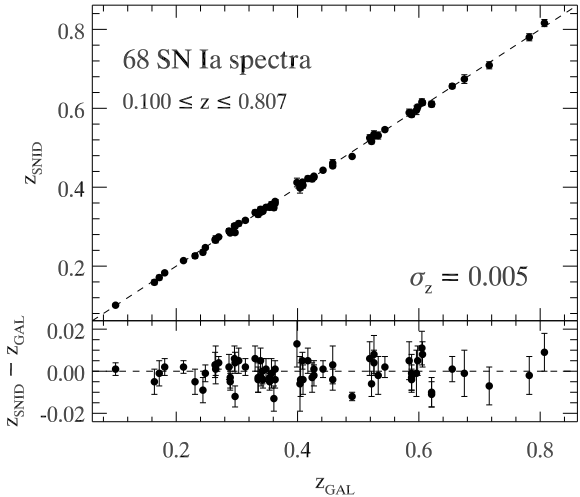}}
\caption{Comparison of redshifts determined from cross-correlations
with SN~Ia spectral templates using SNID ($z_{\rm SNID}$) and from
narrow lines in the host-galaxy spectrum ($z_{\rm GAL}$; {\it upper
panel}). We show the residuals {\it vs.} $z_{\rm GAL}$ in the lower
panel.}\label{f:zsnid}
\end{figure}

We have already shown the excellent agreement between the SNID
correlation redshift and the redshift of the SN host galaxy
(when known) in several publications \citep{Matheson05, Miknaitis07,
Blondin07, Blondin08}.  Figure~\ref{f:zsnid} again shows that the SNID
redshifts agree well with the host-galaxy redshifts, with a dispersion
about the one-to-one correspondence of only \about 0.006.  This figure
contains only the SNe~Ia where we have an independent measurement of
the redshift from a galaxy spectrum.

The template spectra in the SNID database have ages corrected for the
$1/(1+z)$ time-dilation factor expected in an expanding universe
\citep[e.g.,][]{Blondin08}, such that SNID determines ages in the SN
rest frame.  No correction to the age for light-curve width has been
made.  We can then compare these age estimates ($t_{\rm SNID}$) with
those inferred from a well-sampled light curve
\citep[see][]{Miknaitis07}. We expect a one-to-one correspondence
between the light-curve rest-frame age,
\begin{equation}
t_{\rm LC} = \frac{\Delta t_{\rm obs}}{1+z},
\end{equation}
and $t_{\rm SNID}$, where $\Delta t_{\rm obs}$ is the time difference
(in the {\it observer} frame) between maximum light and the time the
spectrum was obtained.

\begin{figure}
\centering
\epsscale{1.1}
\rotatebox{0}{
\plotone{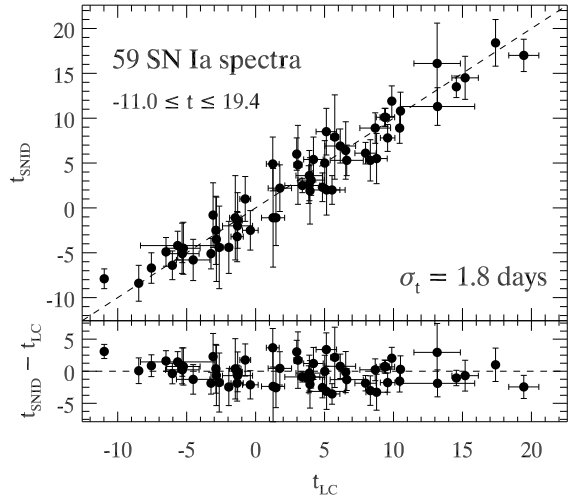}}
\caption{Comparison of SN spectral ages determined using SNID
($t_{\rm SNID}$) and rest-frame light-curve ages ($t_{\rm LC}$) of
high-$z$ SNe~Ia ({\it upper panel}). We show the residuals vs.
$t_{\rm LC}$ in the lower plot.}\label{f:tsnid_mlcs}
\end{figure}

The result is shown in Figure~\ref{f:tsnid_mlcs}.  We use all objects
with a good light-curve fit and a good SNID age, resulting in a total
of 59 SN~Ia spectra with rest-frame light-curve ages in the range $-11.0
\le t_{\rm LC} \le 19.4$~d.  The dispersion about the $t_{\rm SNID}
= t_{\rm LC}$ line is $\sigma_t \approx 2.4$~d, similar to what
was found by \citet{Blondin07}.  We show the residuals versus $t_{\rm
LC}$ in the lower plot.  The mean residual is approximately $-0.7$~d.  
Two points at $t_{\rm LC} < -10$~d are $\sim2\sigma$ off the
null residual line; this is due to the low number of SN~Ia templates
at these early phases.

The fact that the SNID correlation redshift and age measurements agree
so well with the galaxy redshifts and light-curve ages, respectively,
is a strong argument in favor of the similarity of these SNe~Ia with
local counterparts.  This further confirms the results of
\citet{Matheson05}, who found that our classification techniques were
robust and saw no major indications of SN~Ia evolution in our sample.
\citet{Foley08} also see this for the overall sample.

\section{High-Redshift Sample}\label{s:sample}

Over the four seasons, we obtained 329 spectra of 274 objects for a
total of 273 hours of integration.  This has yielded 118 SNe~Ia, 9
SNe~Ia?, 16 SNe~II, 7 SNe~Ib/c, 56 galaxies (which may have some SN
light in the spectrum, but at an undetectable level), 20 AGNs, 4
stars, and 44 unidentified objects.  We provide a detailed assessment
of our object classification in Table~\ref{t:class}.

\begin{deluxetable}{lrrrrr}
\tabletypesize{\scriptsize}
\tablewidth{0pt}
\tablecaption{ESSENCE Spectroscopy Results: The First 4 Years\label{t:class}}
\tablehead{
\colhead{Type} &
\colhead{Year 1} &
\colhead{Year 2} &
\colhead{Year 3} &
\colhead{Year 4} &
\colhead{Total}}

\startdata

Ia        & 15 & 35 & 30 & 38 & 118 \\
Ia?       &  0 &  3 &  2 &  4 &   9 \\
II        &  1 &  2 &  6 &  7 &  16 \\
Ib/c      &  1 &  2 &  2 &  2 &   7 \\
AGN       &  3 &  9 &  0 &  8 &  20 \\
Gal       &  8 & 11 & 20 & 17 &  56 \\
Star      &  2 &  2 &  0 &  0 &   4 \\
Unk/N.S.\ &  6 & 17 & 13 &  8 &  44 \\
Total     & 36 & 81 & 73 & 83 & 274 \\

\tablecomments{Since there were no template observations obtained
before Year 1 and no long baseline observations to easily reject AGNs
and variable stars, there were fewer high-quality SN~Ia candidates in
Year 1.  As a result, Year 1 had fewer SNe~Ia confirmed compared to
later years.}

\enddata

\end{deluxetable}

Objects with no definitive SNID classification or obvious ``by-eye''
classification are labeled as ``Unk.''  Objects with no signal in
their spectra are labeled as ``N.S.''  Together, objects with these
classifications compose the subsample of unidentified objects.

For several objects, we have changed our classification from that published
by \citet{Matheson05}.  This is mostly the result of an improved version
of SNID and additional low-redshift templates for comparison.
Additionally, the \texttt{superfit} routine provided several
classifications that we were not previously able to determine.  In
Table~\ref{t:matheson}, we list the objects with different
classifications between \citet{Matheson05} and this work, as well as
the dominant reason for the reclassification.

\begin{deluxetable*}{lcccl}
\tabletypesize{\scriptsize}
\tablewidth{0pt}
\tablecaption{ESSENCE Object Reclassification \label{t:matheson}}
\tablehead{
\colhead{} &
\colhead{} &
\colhead{\citet{Matheson05}} &
\colhead{The Paper} &
\colhead{} \\
\colhead{ESSENCE ID} &
\colhead{IAUC ID} &
\colhead{Classification} &
\colhead{Classification} &
\colhead{Notes}}

\startdata

c016.wxm1\_04 & \nodata & AGN     & Gal & \\
c022.wxu2\_15 & \nodata & II?     & Ib  & IIb subtype \\
d009.waa6\_16 & \nodata & Gal     & Ia  & \\
d120.wcc1\_2  & \nodata & \nodata & AGN & Not listed by \citet{Matheson05} \\
e141.wdd7\_2  & \nodata & II      & Ib  & IIb subtype \\
e143.wdd7\_3  & \nodata & II      & Ib  & \\
e149.wdd5\_10 & 2003ks  & Ia?     & Ia  & \\
f001.wbb7\_1  & 2003kv  & Unk     & II  & \\
f044.wbb8\_8  & \nodata & Gal     & Ia  & \\
f123.wcc1\_7  & \nodata & Gal     & Ia? & \\
f221.wcc4\_14 & 2003lk  & Ia?     & Gal & \\
f301.wdd6\_1  & \nodata & Ia?     & Ia  & \\
f308.wdd6\_10 & \nodata & Ia?     & Ia  & \\

\enddata

\end{deluxetable*}

\subsection{Survey Efficiency}

In order to meet our goal of \about 200 SNe~Ia over six years, we
tried to use our resources as efficiently as possible to detect and
classify SNe~Ia.  Although the ESSENCE survey has been completed, 
performing an analysis of the efficiency of ESSENCE will provide useful
information for the planning of future surveys (e.g., Pan-STARRS and
DES).

The process of detecting and classifying SNe~Ia can be separated into
two tasks: pre-selecting potential candidates so SNe~Ia represent a
high percentage of spectroscopic targets, and properly classifying the
spectrum of a SN~Ia as a SN~Ia.  The first task is based solely on
parameters derived from imaging.  The second task relies on having a
method of classifying events with high accuracy (such as the SNID
algorithm) and having spectra of sufficient quality for that method to
determine a classification (which depends on observing conditions,
separation from the host galaxy, brightness of the target, exposure
time, the spectrograph's wavelength range, etc.).

\begin{figure}
\centering
\epsscale{0.6}
\rotatebox{90}{
\plotone{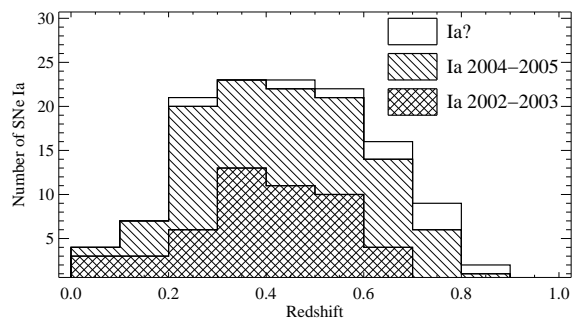}}
\caption{Redshift distribution of spectroscopically identified SNe~Ia
from the first four years of the ESSENCE survey.  The SNe for which we
have a high confidence of being of Type Ia (as determined by SNID) are
plotted in the hashed region.  The open region represents SNe for
which we have less confidence and have been classified
``SNe~Ia?.''}\label{f:zdist}
\end{figure}

\begin{figure}
\centering
\epsscale{0.6}
\rotatebox{90}{
\plotone{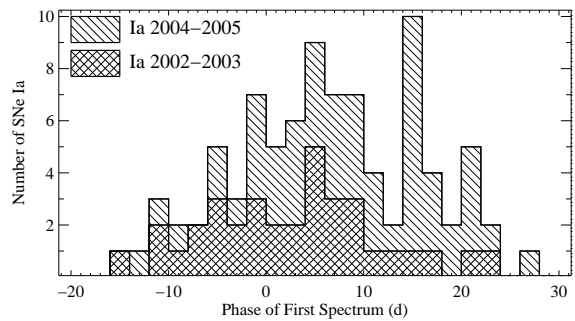}}
\caption{Phase (relative to maximum brightness) distribution of
spectroscopically identified SNe~Ia from the first four years of the
ESSENCE survey and light curves fit by
\citet{Wood-Vasey07}.}\label{f:phase}
\end{figure}

To examine the synergy of our search with the large-aperture telescope
time we used for spectroscopy, we present histograms of our redshift
distribution and phase of the first spectrum of each SN~Ia in
Figures~\ref{f:zdist} and \ref{f:phase}, respectively.  From these
figures we see that as the survey matured, we were able to detect
higher-redshift objects with greater efficiency.  The SNe~Ia detected
during the first two years have an average redshift of 0.38 (and
median redshift of 0.41), while the SNe~Ia detected during years 3 and
4 have an average redshift of 0.45 (and median redshift of
0.42).  We also detected 7 (21) SNe~Ia with $z > 0.7$ ($z > 0.6$) in
the second two years, while we had detected none (4) in the first two
years. Moreover, we were able to detect, observe, and classify SNe~Ia at
later phases after maximum brightness, when the SNe are fainter.  

\begin{figure}
\centering
\epsscale{1.0}
\rotatebox{90}{
\plotone{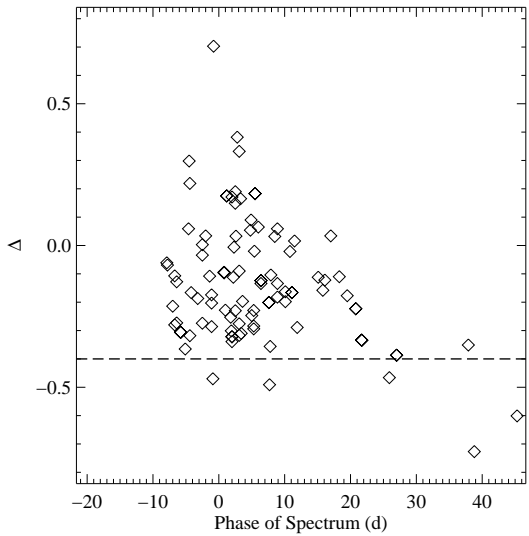}}
\caption{Distribution of measured $\Delta$ of ESSENCE SNe~Ia vs.\ phase
(relative to maximum brightness).  Only objects that had light curves
fit by \citet{Wood-Vasey07} have been plotted.  The dashed line
indicates the $\Delta$ limit beyond which MLCS has to extrapolate to
fit light curves.  As such, objects with $\Delta < -0.4$ should be
discounted.}\label{f:age_delta}
\end{figure}

Figure~\ref{f:age_delta} shows that for objects with larger
$\Delta$\footnote{$M_{V}(t = 0) = -19.504~\textrm{mag} + 0.736\Delta +
0.182 \Delta^{2} + 5 \log_{10} (H_{0}/65)$ \citep{Jha07}.}
(corresponding to being less luminous), we only obtain spectra near
maximum light, while for objects with smaller $\Delta$ (corresponding
to being more luminous), we obtain spectra for a wide range of
phases.  Besides being less luminous, high-$\Delta$ SNe~Ia fade faster
than low-$\Delta$ SNe~Ia, making high-$\Delta$ SNe~Ia become even
fainter relative to low-$\Delta$ SNe~Ia the farther they are from
maximum light.  Both of these biases (faintness and faster declining)
for high-$\Delta$ objects are expected for our sample.  However, one
may argue that since there are fewer total spectra obtained for older
SNe we may be seeing a manifestation of small-number statistics as
opposed to a selection effect.  To test this hypothesis, we performed
a Kolmogorov-Smirnov (K-S) test on the objects with 
$|t| < 5$~d compared to those with $5 <
t < 20$~d and $t > 20$~d; we find the probability that the sets
are chosen from different distributions to be 75.6\% and 99.9\%,
respectively.  The earlier data are marginally consistent, while the
later data are different at a high significance.  This is consistent
with the expected bias.

In Figure~\ref{f:age_delta}, we see that there are 5 SNe (d058,
h311, m022, m043, and m057) with $\Delta < -0.4$, the fiducial
$\Delta$ limit for the multicolor light-curve shape method
\citep[MLCS;][]{Riess96, Jha07} based on the broadest local SN~Ia light
curves.  There are two reasons for such a low value of $\Delta$: the
light curve is broader than any local template, or the light curve is
of low quality and the MLCS fit is not robust.  Of the 5 SNe, three
(h311, m022, m057) fail the $\chi^{2}/\dof < 3$ cutoff set by
\citet{Wood-Vasey07}.  One (m043) has no light-curve points before
maximum, so it is difficult to determine if the light curve is truly
broad.  The final one (d058, with $\Delta = -0.470$) has a
well-sampled light curve and is likely to be truly broad.  Examining
the spectrum, it does not appear that d058 is peculiar, however the
small rest-frame wavelength range (\about3000 -- 5000~\AA) prevents an
investigation of important lines such as \ion{Si}{2} $\lambda$6355.
The first four objects should be ignored since their $\Delta$ values
are likely to be incorrect.  The final object, d058, should be
discounted.  Although $\Delta$ is a well-defined quantity for $\Delta
< -0.4$ (indicating the width of a light curve), for such values we
must extrapolate beyond the local sample, leading to
potentially incorrect distance and luminosity measurements.

\begin{figure}
\centering
\epsscale{0.9}
\rotatebox{90}{
\plotone{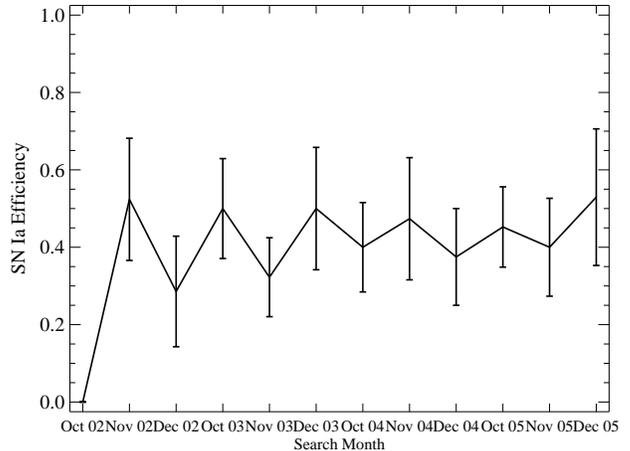}}
\caption{Efficiency of our spectroscopic follow-up observations over
the 12 search months in the first four years of ESSENCE.  The overall
efficiency is 43\%.}\label{f:snepermonth}
\end{figure}

We further investigate if we are limited in our number of classified
SNe~Ia by our search or by our usable time on large-aperture
telescopes.  In Figure~\ref{f:snepermonth}, we show our efficiency
(given by the ratio of SNe~Ia classified to total objects we observed)
as a function of search month.  However, there are many objects where,
after a short exposure, we recognized that it was not a SN~Ia.  An
exposure-time weighted efficiency will be higher than that shown in
Figure~\ref{f:snepermonth}.  Throughout the first four years, we
consistently classified \about 43\% of all spectroscopic targets as
SNe~Ia, perhaps slightly increasing our efficiency with time.  

\begin{figure}
\centering
\epsscale{0.9}
\rotatebox{90}{
\plotone{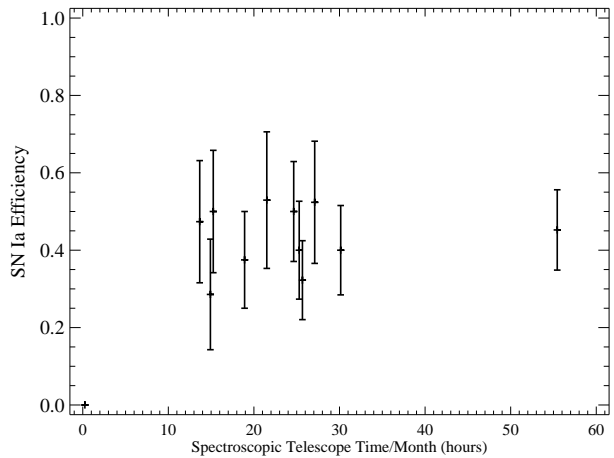}}
\caption{Efficiency of our spectroscopic follow-up observations over
the 12 search months in the first four years of ESSENCE as a function of
total spectroscopic exposure time in a given month.}\label{f:snepertime}
\end{figure}

In Figure~\ref{f:snepertime}, we show that our efficiency does not change
with the amount of spectroscopic time in a given month.  This
indicates that for a broad range of the number of transients detected
per month in our search or for a broad range of spectroscopic
follow-up time in a given month, we would obtain about the same ratio
of SNe~Ia to total spectroscopic targets.

This result is perhaps a little counterintuitive.  There are several
factors involved.  For example, with infinite potential targets from
the search, one expects to be able to properly identify excellent
candidate SNe~Ia, resulting in a higher efficiency.  Similarly, with
infinite spectroscopic follow-up time, one expects to be able to
obtain a spectrum of every transient, resulting in a lower
efficiency.  But there are other factors.  First, when there was a
paucity of potential spectroscopic targets, we would obtain additional
spectra of already confirmed SNe~Ia rather than observe candidates
that had a small possibility of being a SN~Ia.  Another aspect is the
image quality of the search.  If a particular month had poor weather
or bad seeing at Cerro Tololo, we would typically detect brighter,
more isolated objects at lower redshifts.  So even if there were fewer
objects from which to choose, the objects were typically easier to
observe.  It is difficult to disentangle all of these effects, but we
have shown from our measured efficiencies that they naturally balanced
over the first four years of the ESSENCE survey.

\citet{Howell05} showed that the Supernova Legacy Survey (SNLS) was
able to improve their efficiency from 54\% to 71\% by utilizing
multiple colors of the transients and hosts to better predict whether
a transient is a SN~Ia.  The ESSENCE survey, which searched in only
$R$ and $I$, was unable to perform such an analysis.  Furthermore,
ESSENCE only had $R$ and $I$ photometry for the host galaxies,
reducing our ability to determine precise host-galaxy photo-$z$ values.
Obviously, if spectroscopic efficiency is a priority, future surveys
should have many-filter observations of transients and host galaxies.
Additionally, having multiple colors of SNe~Ia will greatly improve
the measurement of SN colors and dust extinction, which in turn should
improve SN distances.

Although we have been able to reduce the number of persistent
transients (AGNs, variable stars) observed spectroscopically, we still
observed 7 AGNs in year 4 (compared to 3, 9, and 0 for years 1, 2, and
3).  These objects were selected for spectroscopic observation despite
some indications that they were AGNs.  This can be attributed to a
human error, and having more humans vet the candidates would help
prevent such future observations.

\begin{figure}
\centering
\epsscale{0.9}
\rotatebox{90}{
\plotone{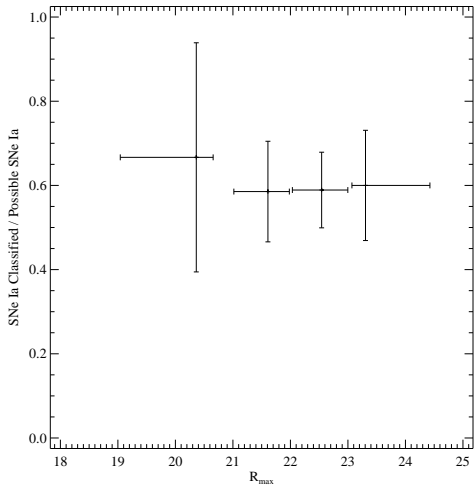}}
\caption{Ratio of the number of SNe~Ia classified to the number of
total objects which may be SNe~Ia (those classified as SNe~Ia,
SNe~Ia?, Gal, Unk, or N.S.) as a function of peak $R$ magnitude.  The
bins are for objects with $R_{\rm max} < 21$, $21 \le R_{\rm max} <
22$, $22 \le R_{\rm max} < 23$, and $R_{\rm max} > 23$ mag, with the
points plotted at the median value for each bin.}\label{f:rmax_eff}
\end{figure}

Finally, if we remove all transients which are definitely not SNe~Ia
(i.e., those objects that have spectra which identify them as other
types of transient objects), we can determine how successful our
spectroscopic classification would have been if we had fewer
contaminants.  Figure~\ref{f:rmax_eff} shows the ratio of the number
of SNe~Ia to the number of total objects which may be SNe~Ia (those
classified as SNe~Ia, SNe~Ia?, Gal, Unk, or N.S.) as a function of
peak $R$ magnitude.  This is a measurement that approximates our
efficiency if we had better photometric preselection.  We do note
that for a fixed amount of galaxy-light contamination it is easier to
classify a bright SN~Ia than a relatively faint SN~Ib/c, indicating
that the ``Gal'' category may contain a lower fraction of SNe~Ia than
the survey as a whole.

Although not all objects classified as Unk, for example, are SNe~Ia,
this ratio still yields a useful approximation of our spectroscopic
efficiency as a function of brightness.  Note that the ratio is a
minimum efficiency since it will increase by either identifying
objects as transients other than SNe~Ia (by decreasing the number of
possible SNe~Ia) or by identifying objects as SNe~Ia (by increasing
the number of SNe~Ia).

Regardless of maximum brightness, we appear to identify similar
percentages of possible SNe~Ia and definite SNe~Ia.  Therefore, we appear to
apply a consistent approach to obtaining sufficient-quality spectra to
identify SNe~Ia for objects of all brightnesses.

\subsection{Sample Demographics}

In the local universe, there are many subtypes of SNe~Ia.  The most
common peculiar subtypes are those of SNe~1991T and 1991bg (described in
Section~\ref{s:classification}), corresponding to more and less
luminous events, respectively.  \citet{Li01:pec} found that 20\% and
16\% of nearby SNe~Ia are similar to SNe~1991T and 1991bg,
respectively.  To determine if the ESSENCE sample of SNe~Ia is
representative of the full SN~Ia population and if the high-redshift
population has different demographics from the low-redshift
population, we examine the peculiarity rate in the ESSENCE sample.

\subsubsection{Peculiar Type Ia Supernova Sample}

SNID cross-correlates a SN spectrum with a library of template SN
spectra.  The template spectra have been classified into subtypes (see
Section~\ref{s:classification}).  If $>50\%$ of the best-fit template
spectra for a given SN are all part of a particular subtype, then the
SN is considered to be of that subtype.

Using SNID, we have determined that five of our SNe~Ia (b004,
d009, d083, d093, and p534) are SN~1991T-like SNe~Ia.
\citet{Matheson05} also claimed that b004 and d083 are similar to
SN~1991T.  The light curves of b004 and d009 were not fit by
\citet{Wood-Vasey07}; however, d083, d093, and p534 were found to have
$\Delta = -0.273$, $-0.365$, and $-0.096$, respectively.  SN~1991T and
the SN~1991T-like SN~1999aa have $\Delta = -0.220$ and $-0.271$,
respectively \citep{Jha07}.

\begin{figure}
\centering
\epsscale{0.6}
\rotatebox{90}{
\plotone{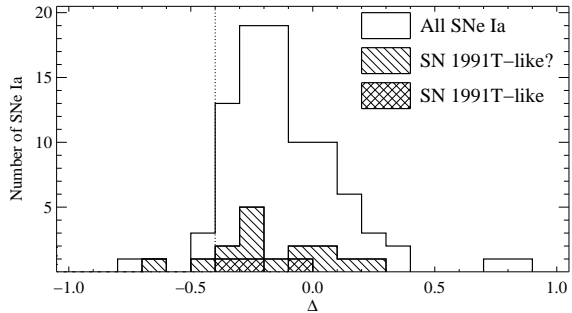}}
\caption{$\Delta$ distribution of spectroscopically identified SNe~Ia
from the first four years of the ESSENCE survey and light curves fit by
\citet{Wood-Vasey07}.  Subsamples of objects identified by SNID as
either being of the subtype consisting of SN~1991T-like objects or
having a best-fit template of that subtype are also shown.  Increasing
$\Delta$ corresponds to narrower light curves and less luminous
SNe~Ia.  The dotted line indicates the $\Delta$ limit beyond which
MLCS has to extrapolate to fit light curves.  As such, objects with
$\Delta < -0.4$ should be discounted.}\label{f:delta91t}
\end{figure}

There are an additional 19 objects which do not pass the threshold
necessary to be safely considered as a SN~1991T-like object, but have
a best-fit template that is of that subtype.  For the sample of 24
SNe~Ia that are either classified as SN~1991T-like or have a best-fit
template of that subtype, there are 16 objects with light curves fit
by \citet{Wood-Vasey07}, having a mean $\Delta$ of $-0.178$.  The
entire ESSENCE sample with light curves fit by \citet{Wood-Vasey07}
has a mean $\Delta$ of $-0.121$.  The $\Delta$ distributions of both
the entire sample and the subsample of SN~1991T-like objects are shown
in Figure~\ref{f:delta91t}.

We have also identified a single object (m226) with a best-fit
template similar to SN~1991bg, but it did not pass the threshold to be
considered SN~1991bg-like.  Since no SN~1991bg-like objects have been
spectroscopically confirmed at high redshift, we pay particular
attention to the classification of this object.  There are several
reasons why we do not believe m226 should be classified as a
SN~1991bg-like object.  First, the spectrum only matched three SNID
template spectra, two of which were SNe~Ia-norm.  Second, the
best-match template spectrum is of SN~1999gh at 42~d past maximum
light, while the light-curve age is 24~d past maximum light.  The
best-match template with a spectral age near that of the light-curve
age of m226 is SN~2002bo (at 29~d past maximum light), which is a
SN~Ia-norm.  Third, SN~1999gh is a low-luminosity SN~Ia with a deep
\ion{Si}{2} $\lambda$5800 line; however, it does not share many of the
spectral features of SN~1991bg \citep{Matheson08}.  Fourth, the
spectrum of m226 has a relatively low S/N ratio, where proper
subtyping is dubious.  Finally, the light-curve fit of m226 yielded
$\Delta = -0.227$, which is on the opposite end of the $\Delta$ range
from SN~1991bg-like objects.  With all of these data, we are therefore
unable to properly classify the subtype of m226.

\subsubsection{Peculiarity Rate}

The spectral features that distinguish SNe~1991T and 1991bg
from SNe~Ia-norm can be hidden by low-S/N spectra, spectra
with a restricted rest-frame wavelength range, and spectra obtained
well after maximum brightness.  Nevertheless, we still have observed
very few peculiar SNe~Ia in the ESSENCE survey.  Of the 131 SNe~Ia and
SNe~Ia?, we have spectroscopically confirmed 5 SNe~Ia similar to
SN~1991T (with 19 additional potential objects).  Therefore, we have
determined that 4--19\% of the SNe~Ia from the ESSENCE survey are
similar to SN~1991T, while no ESSENCE SN is similar to SN~1991bg.
Because of selection effects (discussed below), our upper limit of
SN~1991T-like objects may be lower than the real value; however, since
the lower limit is determined by definitive detections, it cannot be
smaller.  \citet{Bronder08} spectroscopically confirmed 2--3
high-redshift SNLS SNe~Ia out of a sample of 54 to be comparable to
SN~1991T; this rate is similar to that found in the ESSENCE survey.

Comparing to low-redshift SN searches, \citet{Li01:pec} found that
20\% and 16\% of nearby SNe~Ia are similar to SNe~1991T and 1991bg,
respectively.  Our raw rate of SN~1991bg-like objects appears to be
very low compared to the low-redshift rates.  However, with a typical
limiting magnitude of $R = 24$~mag, we would only expect to be
complete for SN~1991bg-like objects out to $z \approx 0.35$.  We have
detected 46 SNe~Ia (or SNe~Ia?) with $z \le 0.35$.  If the
low-redshift rates are similar to the rates at $z \approx 0.35$, we
would expect to have detected 7 SN~1991bg-like SNe~Ia in this
subsample.  If there are no SN~1991bg-like SNe~Ia at high redshift,
but the ratio of SN~1991T-like to SNe~Ia-norm remains constant, then
one would expect 24\% of high-redshift SNe~Ia to be similar to
SN~1991T.  Considering the selection effects involved, we believe that
the rate of SN~1991T-like SNe~Ia spectroscopically confirmed in the
ESSENCE survey is consistent with the rate found at low redshift.
Conversely, we find that the raw rate of SN~1991bg-like SNe~Ia found
in the ESSENCE survey is low.  A further indication of a low raw rate
is that no survey has spectroscopically confirmed a SN~1991bg-like
SNe~Ia at high redshift.

\begin{figure}
\centering
\epsscale{1.1}
\rotatebox{90}{
\plotone{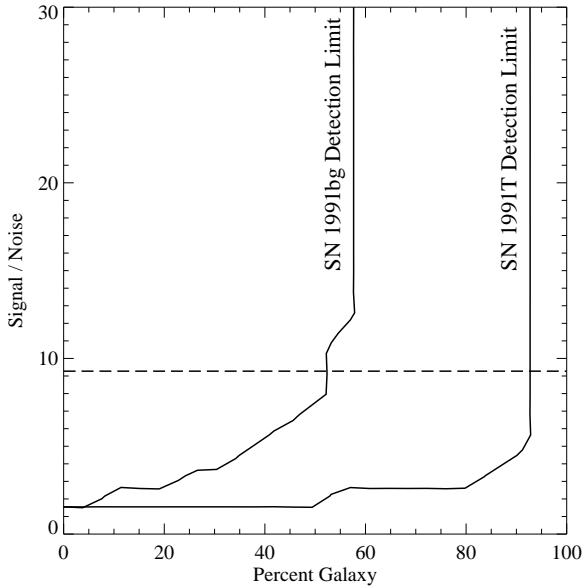}}
\caption{Contours representing the region of S/N-galaxy contamination
parameter space that SNe~Ia similar to SNe~1991T and 1991bg at $z =
0.35$ will have a correct subtype classification from SNID.  We have
made no assumption about the relative luminosities of the SNe or host
galaxies.  If a spectrum is to the upper-left of the contour, then it
is correctly classified into its subtype.  The dashed line is the
median S/N for our sample of SNe~Ia with $z \le 0.35$.  The detection
efficiencies show that we should recover nearly every SN~1991T-like
SN~Ia to this redshift, while the detection of SN~1991bg-like objects
is heavily dependent on the amount of galaxy
contamination.}\label{f:cont_91bg}
\end{figure}

To determine a reasonable detection efficiency for SNe similar to
SN~1991bg, we performed a basic simulation.  Starting with the
maximum-light spectrum of SN~1991bg, we added a varying amount of
noise and galaxy contamination (using the spectrum of an elliptical
galaxy).  We then processed the spectra with SNID, removing SN~1991bg
from our sample of template SNe and restricting our wavelength range
to 3000--6600~\AA, an appropriate rest-frame wavelength for $z =
0.35$.  From this, we were able to determine what parameters of S/N
and galaxy contamination would yield a correct subtype classification
by SNID.  We repeated this process for SN~1991T with an Sb galaxy.
The results are shown in Figure~\ref{f:cont_91bg}.  For this
simulation, we did not make any assumption about the SN or host galaxy
luminosities.  SN~1991bg-like objects should be less luminous and have
more luminous host galaxies than SN~1991T-like objects.  This would
mean that the average spectrum of a SN~1991bg-like object should have
more galaxy contamination than that of a SN~1991T-like object.

From Figure~\ref{f:cont_91bg}, we see that for $z \le 0.35$, we should
detect essentially all SN~1991T-like objects.  However, for
SN~1991bg-like objects, the detection efficiency is strongly dependent
on galaxy contamination.  We would detect most SN~1991bg-like objects
with $< 60$\% galaxy contamination.

If we assume that the low- and high-redshift peculiarity rates are the
same, then selection effects and classification systematics must
decrease the SN~1991T-like classification rate by a factor of 1--5.
Considering that SN~1991bg-like objects tend to be found in high
surface-brightness galaxies, many of these objects will be fainter
than their galaxies at their position.  We are able to remove a
significant amount of galaxy light during spectral reductions
(particularly for elliptical galaxies where the radial light
distribution is smooth), but it is still likely that a significant
number of spectra of these objects would be dominated by galaxy
light, and therefore would not be classified in this subtype by SNID.

If we assume that most spectra of SN~1991bg-like objects will have
galaxy contaminations of 25--75\% and a S/N of 4--15 (similar to most
of our spectra), we can determine the percentage of these spectra that
SNID would classify as SN~1991bg-like objects.  Placing these
constraints in the parameter space shown in Figure~\ref{f:cont_91bg},
we see that 54\% of the remaining parameter space results in a
positive detection.

If we assume that the selection effects and classification systematics
are essentially equal to the detection efficiency of SNID given our
assumed parameters above, we require a bias factor of 1.86 for the
SN~1991bg-like objects compared to the SN~1991T-like objects.  Using
this factor, we would expect the SN~1991bg-like peculiarity rate to be
2--8\% for the ESSENCE sample.  This corresponds to 1--4 SNe in our $z
\le 0.35$ sample, consistent with our non-detection of SN~1991bg-like
events in our sample.  Increasing the sample size and constraining the
bias factors will greatly improve the constraints.

A secondary method to examine the rate of SN~1991bg-like objects is by
measuring the light-curve shapes of our SNe~Ia.  SN~1991bg-like
objects have significantly different light-curve shapes than normal
SNe~Ia, with corresponding $\Delta > 1$.  Our sample contains 31
SNe~Ia with $z \le 0.35$ and $\Delta$ information.  If the
local rate of SN~1991bg-like objects is constant to $z = 0.35$, then
we would expect to have 5 SN~1991bg-like objects in this sample.  None
of the SNe~Ia in this sample have $\Delta > 1$.  Assuming that all
objects spectroscopically similar to SN~1991bg also have $\Delta > 1$,
the nondetection of any of these objects is $2.2\sigma$ from the
expected value.


A significant bias against classifying SN~1991T-like SNe~Ia
(particularly objects similar to SN~1999aa) at high redshift is
that their post-maximum spectra are very similar to those of
SNe~Ia-norm.  For instance, there is essentially no difference between
the spectra of the normal SN~1994D and SN~1999aa one week after
maximum brightness \citep{Garavini04}.  Therefore, for a large
percentage of our SNe~Ia, we have no real constraint on the
peculiarity of the individual object.  More detailed modeling is
necessary to determine the classification success as a function of
phase, S/N, and wavelength range of a particular spectrum.

\section{Conclusions}\label{s:conclusions}

We have presented optical spectra of targets selected for follow-up
observations during the first four years of the ESSENCE survey.  We have
shown that as the survey matured, we were able to detect more SNe~Ia
at higher redshifts and at later phases without reducing efficiency
for a total of 127 SNe~Ia and possible SNe~Ia (``Ia?'').  There were
more SNe~Ia observed in the second two years than in the first two
years, showing that we are on track to reach the goal of the ESSENCE
survey: constraining the equation-of-state parameter of dark energy to
$\lesssim 10$\% by observing \about 200 SNe~Ia over the redshift range
$0.2 \lesssim z \lesssim 0.8$.

The spectra of the ESSENCE high-redshift SNe~Ia are broadly consistent
with those of their low-redshift counterparts; see \citet{Foley08} for
a detailed comparison of composite spectra made from this sample.  If
the ESSENCE SNe~Ia were drastically different from their local
counterparts, then the correlations of redshift and age found by SNID
would have larger scatter.  We are able to automatically classify the
majority of our objects using SNID.  However, human intervention,
through visual inspection or using a second fitting program, was
necessary to properly classify several objects.  SNID is able to
accurately quantify the redshift and phase determination for our
sample of SNe~Ia.  The peculiarity rate of the spectroscopically
distinct SN~1991T-like and SN~1991bg-like objects are consistent with
low-redshift rates.  None of the spectra from our sample or light
curves fit by \citet{Wood-Vasey07} are similar to the
spectroscopically peculiar SN~1991bg.  However, if we consider
systematic effects, these non-detections are consistent with
low-redshift rates.

\begin{acknowledgments} 
This research is based in part on observations obtained at the Cerro
Tololo Inter-American Observatory, which is operated by the
Association of Universities for Research in Astronomy, Inc. (AURA)
under cooperative agreement with the National Science Foundation
(NSF); the European Southern Observatory, Chile (ESO Programmes
170.A-0519 and 176.A-0319); the Gemini Observatory, which is operated
by the Association of Universities for Research in Astronomy, Inc.,
under a cooperative agreement with the NSF on behalf of the Gemini
partnership: the NSF (United States), the Science and Technology
Facilities Council (United Kingdom), the National Research Council
(Canada), CONICYT (Chile), the Australian Research Council
(Australia), CNPq (Brazil), and CONICET (Argentina) (Programs
GN-2002B-Q-14, GS-2003B-Q-11, GN-2003B-Q-14, GS-2004B-Q-4,
GN-2004B-Q-6, GS-2005B-Q-31, GN-2005B-Q-35); the Magellan Telescopes
at Las Campanas Observatory; the MMT Observatory, a joint facility of
the Smithsonian Institution and the University of Arizona; and the
F.~L. Whipple Observatory, which is operated by the Smithsonian
Astrophysical Observatory. Some of the data presented herein were
obtained at the W.~M. Keck Observatory, which is operated as a
scientific partnership among the California Institute of Technology,
the University of California, and the National Aeronautics and Space
Administration (NASA); the Observatory was made possible by the
generous financial support of the W.~M. Keck Foundation.

The ESSENCE survey team is very grateful to the scientific and
technical staff at the observatories we have been privileged to use.

{\it Facilities:} 
  \facility{Blanco (MOSAIC II)}, 
  \facility{CTIO:0.9m (CFCCD)}, 
  \facility{Gemini:South (GMOS)}, 
  \facility{Gemini:North (GMOS)}, 
  \facility{Keck:I (LRIS)},
  \facility{Keck:II (DEIMOS, ESI)},
  \facility{ESO VLT (FORS1)},
  \facility{Magellan:Baade (IMACS)}, 
  \facility{Magellan:Clay (LDSS2)},
  \facility{Magellan:Clay (LDSS3)}.

The ESSENCE survey is supported by the US National Science Foundation
through grants AST--0443378 and AST--0507475. The Dark Cosmology Centre
is funded by the Danish National Research Foundation.
A.C. acknowledges support from grants FONDECYT 1051061, FONDAP
15010003, and P06-045-F (Millennium Center for Supernova Science
funded by programs Bicentenario de Ciencia y Tecnologia de CONICYT and
ICM de MIDEPLAN).  A.V.F.'s supernova group at U.C. Berkeley has been
supported by many NSF grants over the past two decades, most recently
AST--0307894 and AST--0607485.  P.M.G. is supported in part by NASA
Long-Term Astrophysics Grant NAG5-9364 and NASA/{\it HST} Grant
GO-09860.  S.J. thanks the Stanford Linear Accelerator Center for
support via a Panofsky Fellowship.  R.P.K. enjoys support from
AST--0606772 and PHY--9907949 to the Kavli Institute for Theoretical
Physics.  G.P. acknowledges support by the Proyecto FONDECYT 3070034.
A.R. thanks the NOAO Goldberg fellowship program for its support.

Our project was made possible by the survey program administered by
NOAO, and builds upon the data-reduction pipeline developed by the
SuperMacho collaboration.  
\end{acknowledgments}

\bibliographystyle{apj}
\bibliography{astro_refs}

\clearpage
\LongTables
\begin{landscape}
\begin{deluxetable}{llllllllllllll}
\tabletypesize{\scriptsize}
\tablenum{2}
\tablewidth{0pt}
\tablecaption{ESSENCE SPECTROSCOPIC TARGETS\label{t:sntable}}
\tablehead{
\colhead{ESSENCE ID\tablenotemark{a}} &
\colhead{IAUC ID\tablenotemark{b}} &
\colhead{UT Date} &
\colhead{Telescope} &
\colhead{Type\tablenotemark{c}}  &
\colhead{Subtype} &
\colhead{$z$} &
\colhead{$z$} &
\colhead{Template} &
\colhead{Phase} &
\colhead{Phase} &
\colhead{$\Delta$} &
\colhead{Disc.} & 
\colhead{Exp.} \\
\colhead{} &
\colhead{} &
\colhead{} &
\colhead{} &
\colhead{} &
\colhead{} &
\colhead{(Gal)\tablenotemark{d}} &
\colhead{(SNID)\tablenotemark{e}} &
\colhead{} &
\colhead{(SNID) (d)} &
\colhead{(LC) (d)} &
\colhead{} &
\colhead{Mag.} & 
\colhead{(s)}
}

\startdata

a002.wxc1\_04                      & \nodata & 2002-12-06.03  & VLT        & Gal     & \nodata  & 0.315         & \nodata       & \nodata & \nodata      & \nodata       & \nodata  & \nodata &  900 \\
b001.wxc1\_14                      & \nodata & 2002-11-03.23  & MMT        & Unk     & \nodata  & \nodata       & \nodata       & \nodata & \nodata      & \nodata       & \nodata  & \nodata &  1800 \\
b001.wxc1\_14                      & \nodata & 2002-11-06.27  & KII/ESI    & Unk     & \nodata  & \nodata       & \nodata       & \nodata & \nodata      & \nodata       & \nodata  & \nodata &  1600 \\
b001.wxc1\_14                      & \nodata & 2002-11-11.32  & KI/LRIS    & Unk     & \nodata  & \nodata       & \nodata       & \nodata & \nodata      & \nodata       & \nodata  & \nodata &  1800 \\
b001.wxc1\_14                      & \nodata & 2002-12-05.26  & GMOS       & Unk     & \nodata  & \nodata       & \nodata       & \nodata & \nodata      & \nodata       & \nodata  & \nodata &  2x1800 \\
b002.wxh1\_01                      & \nodata & 2002-11-01.44  & KII/ESI    & Star    & \nodata  & \nodata       & \nodata       & \nodata & \nodata      & \nodata       & \nodata  & \nodata &  900 \\
b003.wxh1\_14                      & 2002iu  & 2002-11-01.43  & KII/ESI    & Ia      & norm     & \nodata       & 0.115 (0.005) & 1999ee  & $-3.8$ (3.2)      & \nodata       & \nodata  & \nodata &  600 \\
b004.wxt2\_06                      & 2002iv  & 2002-11-02.45  & KII/ESI    & Ia      & 91T      & 0.231         & 0.226 (0.006) & 2005eq  & $-4.9$ (1.6)      & \nodata       & \nodata  & \nodata &  1200 \\
\enddata

\tablenotetext{a}{ESSENCE internal identification.  The first letter
  indicates the month in the observing season.  This is followed by a
  sequential number as targets are discovered.  The remaining letters
  and numbers show the specific ESSENCE field where the object was located.}

\tablenotetext{b}{Note that not all objects judged to be SNe have
  official International Astronomical Union names.}

\tablenotetext{c}{Our best guess as to classification of the object.
  Ia? indicates a lack of certainty in the identification as a SN~Ia.
  Objects marked ``Unk'' are unknown.  N.S. indicates that the
  telescope was pointed to the object, but no exposure was taken or
  the exposure contained no signal.}

\tablenotetext{d}{Redshift measured from narrow emission or absorption
  lines from the host galaxy. All galactic redshift errors are $<0.001$.}

\tablenotetext{e}{Redshift measured from the SN spectrum by SNID.}

\end{deluxetable}
\clearpage
\end{landscape}


\end{document}